\documentstyle[preprint,aps]{revtex}
\date{\today}
\begin{document}
\draft
\title{Lower Bound for the Fermi Level Density of States of a 
Disordered D-Wave Superconductor in Two Dimensions}
\author{
K. Ziegler$^{a,b}$, M.H. Hettler$^{c,d}$, and P.J. Hirschfeld$^c$
}
\address{$^a$Max-Planck-Institut f\"ur Physik Komplexer Systeme,
Au\ss enstelle Stuttgart, Postfach 800665, D-70506 Stuttgart, Germany\\
$^b$ Institut f\"ur Physik, Universit\"at Augsburg, D-86135 Augsburg,
Germany\\
$^c$ Dept. of Physics, Univ. of Florida, Gainesville, FL 32611, USA\\
$^d$ Dept. of Physics, Mail Location 11, University of Cincinnati, 
Cincinnati, OH 45221, USA}
\maketitle 
\begin{abstract}
We consider a disordered d--wave superconductor in two dimensions. 
Recently, we have shown in an exact calculation that for a lattice model
with a Lorentzian 
distributed random chemical potential the quasiparticle 
density of states at the 
Fermi level is nonzero. As the exact result holds only
for the special choice of the Lorentzian, we employ different methods
to show that for a large class of distributions, including the Gaussian
distribution, one can establish a nonzero lower bound for the Fermi level
density of states. 
The fact that the tails of the distributions are unimportant in
deriving the lower bound shows that the exact result obtained 
before is generic.\end{abstract}

\section{Introduction}

Considerable evidence for d-wave superconductivity in the 
high-temperature cuprate superconductors has led to interest in studying 
the effect of disorder on  d-wave paired systems. Unlike s-wave 
superconductors, where Anderson's theorem \cite{Anderson}
predicts negligible effect of
nonmagnetic impurities on thermodynamic properties, simple
defects are expected to be pairbreaking in superconductors with
gap nodes, and are in fact generally thought to induce finite density of
quasiparticle states $N(0)$ at the Fermi level. As in disordered normal
metals, one might expect properties of such systems to depend 
strongly on dimensionality. In fact, Nersesyan et al. (NTW)
have shown\cite{ntw} that the usual t-matrix approximation for impurity
scattering, which is exact in the dilute
limit in 3D, breaks down for a strictly 2D d-wave superconductor.
By mapping the problem onto a continuum model of Dirac fermions in a random 
gauge field, subsequently solved by bosonization methods, NTW claimed
that the density of states of such systems must go to zero at the Fermi
level as a power law, $N(E)\propto E^\alpha$.  
Later it was realized that for a realistic d--wave SC with 4 nodes on the
Fermi surface their result might not be applicable \cite{ntw1}.
Although the real materials in question are quasi-2D,
it is of considerable importance to establish
the effect of  disorder  in the strictly 2D case because the existence
of a 2D-3D crossover at low energies could invalidate the standard picture
of low-temperature thermodynamics in a d-wave superconductor developed
under the assumption of a finite residual density of states $N(0)$.\\

Recently \cite{zhh}, we have shown that for a lattice model of a
disordered d--wave
superconductor (SC) in two spatial dimensions, one can obtain 
an exact result for
the density of states (DOS) $N(E)$, provided that the disorder is modeled by
a Lorentzian distribution of the chemical potential. The result was
a finite DOS at the Fermi level, $N(0)/N_o \propto \gamma \log 4\Delta_0/
\gamma$
with $N_o$ the normal DOS at the Fermi level, $\Delta_0$ the maximum value of
the superconducting order parameter over a circular Fermi surface 
and $\gamma$ the width of the Lorentzian
distribution. We also quoted rigorous lower bounds for $N(0)$ for
a large class of disorder distributions which
we obtained using methods developed in a different context.
One might worry that our result for a Lorentzian distribution, 
while simple to obtain and exact for all energies, could be nongeneric, in
the sense that a perturbation series based on a Lorentzian distribution
cannot be defined due to the divergence of all moments. The proof of
our lower bounds for the DOS in the case of more general disorder
distributions then acquires a special importance.\\

In this paper we therefore present
in some detail the derivation of the nonzero lower bound
for the DOS at the Fermi level
which in a different context was first given in Ref. \cite{zie1}.
We stress that since our results are lower bounds, no arguments about the 
dependence of the DOS on disorder strength can be made.  It suffices
for our purposes to show that a lower bound exists, and
that its existence does not depend on the specifics of the tails of the
distribution
i.e. power law decay, exponential decay or compact support of the
distribution will all give a nonzero lower bound for the DOS.

\vskip .2cm

The paper is organized as follows: First, we formulate the problem
and give a general outline of the proof. Second, we show preliminary
calculations which will be used in the proof. We then derive the
nonzero lower bound for a certain class of Hamiltonians. Finally,
we show that the Hamiltonian of interest belongs to this class. We conclude
with final remarks about cases where the method fails to give a nonzero
lower bound (e.g. s-wave superconductors).

\section{Formulation of the problem and outline of the proof}
The problem is defined\cite{zhh} by the Bogliubov-- de Gennes Hamiltonian
\begin{equation}
H=-(\nabla^2+\mu)\sigma_3+{\hat\Delta}^d\sigma_1 , 
\label{hamil}
\end{equation}
$\mu$ being the random chemical
potential, distributed according to a symmetric 
distribution function $P(\mu_r)$
($\sigma_i$ are the Pauli matrices, $\sigma_0$ is the $2\times2$ 
identity matrix). The kinetic energy operator $-\nabla^2$ is taken to
act as $\nabla^2\Psi(r)=\Psi(r+2e_1)+\Psi(r-2e_1)+\Psi(r+2e_2)
+\Psi(r-2e_2)$ on  a function $\Psi(r)$ of the
sites $r$ of a 2D square lattice $\Lambda$ spanned by the unit
vectors $e_1$ and $e_2$.  Note that this function involves displacements
of two lattice sites rather than one, as would be the case
in the simplest  tight-binding representation of the lattice kinetic energy.
For a system of fermions in the thermodynamic limit, the bare kinetic
energy will then have a band representation quite similar to
the usual tight-binding form, with no particular distinguishing features
near the Fermi level.
In the above definition of the kinetic energy, we have taken 
the hopping matrix element as our unit of energy.
The bilocal lattice operator ${\hat \Delta}^d\equiv\Delta_{r,r^\prime}$ is
taken be ${\hat \Delta}^{d}\Psi(r)=\Delta
[\Psi(r+e_1)+\Psi(r-e_1)\pm\Psi(r+e_2)\pm\Psi(r-e_2)]$.
The Matsubara Green function $G(iE)=(iE\sigma_0 - H)^{-1}$
determines the density of states in the usual way, namely by 
\begin{equation}
N (E) = \frac{-1}{2 \pi} Im Tr_2 \langle G_{r,r}
(iE\rightarrow E + i\epsilon)\rangle
\end{equation}
where the trace $Tr_2$ refers to the $2\times2$ structure of the Hamiltonian,
corresponding to
quasiparticles and quasiholes of the superconductor.
$\langle \ldots \rangle$ denotes the disorder average, which consists 
of integration over the disorder variable (the chemical potential) 
at every site of the lattice, with a measure given by $P(\mu_r) d\mu_r$.\\

To derive a lower bound of the average DOS we first write the Green
function as 
\begin{eqnarray}
G(iE) = \frac{i (iE\sigma_0 + H)}{2 E} & \left( (\tilde{H}-iE\sigma_0)^{-1}
- (\tilde{H}+iE\sigma_0)^{-1} \right)
\label{gfun1}
\end{eqnarray}
where the new Hamiltonian is ${\tilde H}=H D\sigma_3=
-(\nabla^2+\mu)D\sigma_0-i\Delta D\sigma_2$ has been
introduced for formal reasons (cf. Ref. \cite{zhh} and below). 
The matrix $D$ is
diagonal with matrix elements $D_{r,r'} = (-1)^{r_1+r_2}\delta_{r,r'}$
($r_1, r_2 $ are the two components of the 2D $r$--vector).
Eq. (\ref{gfun1}) holds for any distribution of randomness in the
Hamiltonian, before averaging. We are interested in the DOS at the Fermi
level, i.e. at zero energy. This means that after the analytic continuation
$iE\rightarrow E+ i\epsilon$ we will set $E=0$. Consequently, 
$iE\rightarrow i\epsilon$ (with positive $\epsilon$), and the
local DOS at the Fermi level at lattice site $r$ reads
\begin{equation}
N_r(0) ={-1\over2\pi}Tr_2Im G_{rr}(i\epsilon)
={-i\over4\pi}Tr_2[({\tilde H}-i\epsilon\sigma_0)^{-1}_{rr}
-({\tilde H}+i\epsilon\sigma_0)^{-1}_{rr}]
={\epsilon\over2\pi} Tr_2({\tilde H}^2+\epsilon^2\sigma_0)^{-1}_{rr}.
\label{localdos}
\end{equation}
From the analytic properties
of $G$ it follows that $N_r$ is nonnegative (either positive or zero).\\

The average DOS $(1/|\Lambda|) \langle \sum_{r \in \Lambda} N_r \rangle$ can
be estimated from below using the method worked out in Ref. \cite{zie1}.  
The central idea of the proof is to
divide the lattice $\Lambda$ into finite lattice blocks $\{ S_j\}$. We then 
evaluate the average DOS on these lattice blocks and also the
contribution from the interactions of the lattice blocks.
On the lattice blocks a ``coarse graining'' method will be used
by relating the disorder integration over all other sites 
to one at the ``center'' of $S$. On this ``center'' 
site the range of integration
of the random variable $z_r= \mu_r D$ will be restrained to a finite
interval $[-a,a]$. The cutoff $a$ eliminates the contribution of the
tails of the distribution. Since $N_r$ is nonnegative, the tail
contribution can only add to the result obtained by integrating over
$[-a,a]$. Thus, {\it if} we are able to find a {\it nonzero}
average DOS by integrating only
over the finite interval $[-a,a]$ we have obtained a nonzero DOS without
relying on tail contributions. This explains why distributions with
power law tails (e.g. 
Lorentzian distribution) lead qualitatively to the same results 
as, for example, the Gaussian distribution or distributions with exponential
decay.\\

The proof rests on an identity (Eq. (14) in the  next section)
that is intimately connected with the fact that the local DOS $N_r$
(before averaging)
is nonnegative. It also relies on the infimum of the disorder 
distribution in the restricted range $[-a,a]$  being finite. This puts 
some limits on the applicability of the proof to compact distributions, but
it always holds for unbounded distributions like the Gaussian or 
the Lorentzian. 
The result can be summarized by the statement:
{\it For any finite subregion $S$ of the lattice $\Lambda$ with boundary
$\partial S$, defined by the lattice sites of $\Lambda\backslash S$ which
are connected
to $S$ by the matrix elements of ${\tilde H}$, there exists a 
distribution dependent positive
constant $P_S$, related to a restricted disorder distribution on $S$, with}
\begin{equation}
{1\over|\Lambda|}\Big\langle
\sum_{r\in \Lambda}N_r\Big\rangle \ge P_S(1-|\partial S|/|S|).
\label{lowbound0}
\end{equation}
Since the block size $|S|$ grows faster than the size of the boundary
$|\partial S|$ the right hand side is positive above a certain
block size.\\

It should be noted  that the method of this paper
will  not give a {\it nonzero} lower
bound for the DOS for {\it every} Hamiltonian. 
We will determine  the conditions for the lower bound
to be nonzero and show that the Hamiltonian of interest 
(d--wave SC) fulfills these conditions. We will also
show that for an isotropic s--wave SC with a local order parameter
the method will only yield a (trivial) vanishing  bound for the 
Fermi level DOS.
\section{Preliminary Considerations}
As a first step we evaluate the integral $\int_{-\infty}^\infty N_r dz_r$.
For this purpose the identity
\begin{equation}
I_S({\tilde H}+I_SAI_S)^{-1}I_S=\Big[({\tilde H}^{-1})_S^{-1}
+I_SAI_S\Big]^{-1}_S
\label{projected}
\end{equation}
is useful. 
$I_S$ is
the projector onto the region $S$, and $(...)^{-1}_S=I_S(I_S...I_S)^{-1}I_S$
is the inverse on the region $S$. 
The proof of this identity is given in Appendix A.
Choosing $S=\{ r\}$, i.e., just a single lattice site, we note that the
$2\times2$ matrix $(({\tilde H}+z_r\sigma_0+i\epsilon
\sigma_0)^{-1})^{-1}_{\{ r\}}$ is diagonal. Furthermore, it is 
proportional to the unit matrix $\sigma_0$ as a 
consequence of the definition of ${\tilde H}$, which involves only 
$\sigma_0$ and $\sigma_2$. (Terms of the inverse that are
proportional to $\sigma_2$ are nonlocal and, consequently,
projected out by $I_{\{ r\}}$). Therefore, we can write
\begin{equation}
(({\tilde H}+z_r\sigma_0+i\epsilon\sigma_0)^{-1})^{-1}_{\{ r\}}=\pmatrix{
X_{r}+iY_{r}&0\cr
0&X_{r}+iY_{r}\cr
},
\label{diagcond}
\end{equation}
where $Y_{r}\propto \epsilon >0$. 
For the special choice $I_{\{ r\}}AI_{\{ r\}}=-z_r\sigma_0$
in Eq. (\ref{projected}) we obtain
\begin{equation}
({\tilde H}+i\epsilon\sigma_0)^{-1}_{rr}=
\big((({\tilde H}+z_r\sigma_0+i\epsilon\sigma_0)^{-1})^{-1}_{\{ r\}}
-z_r\sigma_0\big)^{-1}_{rr}.
\end{equation}
This gives for the local DOS of Eq.
(\ref{localdos})
\begin{equation}
{1\over\pi}  Y_{r}[(X_{r}-z_r)^2+Y_{r}^2]^{-1}.
\end{equation}
The integration over $z_r$ leads to
\begin{equation}
\int_{-\infty}^\infty N_r dz_r=1.
\label{intrho}
\end{equation}
This result will be used below. It is nontrivial as, for example,
for an s--wave SC with a {\it local} order parameter term $\Delta \sigma_1$
the equivalent of Eq. (\ref{diagcond}) would have also off--diagonal
entries. In fact, we have shown in \cite{zhh} that 
the corresponding expression of the local DOS of the s-wave superconductor reads
\begin{equation}
N_r(0) = -{1\over\pi}{i\epsilon\over2\sqrt{\Delta^2+\epsilon^2}}
[(-\nabla^2-\mu-i\sqrt{\Delta^2+\epsilon^2})^{-1}_{rr}
-(-\nabla^2-\mu+i\sqrt{\Delta^2+\epsilon^2})^{-1}_{rr}],
\label{gswave1}
\end{equation}
i.e., it is proportional to $\epsilon/\sqrt{\Delta^2+\epsilon^2}$.
This implies a vanishing DOS in the limit 
$\epsilon\to0$ as long as the superconducting order parameter is nonzero.
Of course, that is what is expected for a SC with a nonvanishing gap
everywhere on the Fermi surface (Anderson's theorem) \cite{Anderson}.\\

The local DOS Eq. (\ref{localdos}) can be written in a differential form as
\begin{eqnarray}
N_r={i\over4\pi}{\partial\over\partial z_r}
[\log\det({\tilde H}-i\epsilon\sigma_0)-\log\det({\tilde H}+i\epsilon\sigma_0)]
\nonumber\\
={i\over4\pi}{\partial\over\partial z_r}
[\log\det({\bf 1}-2i\epsilon({\tilde H}+i\epsilon\sigma_0)^{-1})],
\label{localdos1}
\end{eqnarray}
which follows from 
the fact that the differentiation with respect to $z_r$ picks the 
$r, r$--component of ${\tilde H} \pm i\epsilon\sigma_0$.
We define a matrix $A$ as
\begin{equation}
A:=-2i\epsilon({\tilde H}+i\epsilon\sigma_0)^{-1}.
\end{equation}
Because the DOS is nonnegative the function $i\log\det({\bf 1}+A)$ is a
nondecreasing function of $z_r$. As a consequence of the Eq. (\ref{intrho})
we have the following integral
\begin{equation}
\int_{-\infty}^\infty N_rdz_r
={i\over4\pi}\log\det({\bf 1}+A)\Big|_{z_r=-\infty}^{z_r=\infty}=1.
\label{integral}
\end{equation}
\section{Lower bound of the average DOS}

\subsection{Coarse Graining and Elimination
of the Distribution Tails}
Now we consider the  average local DOS summed over all lattice sites  
on a finite lattice block $S$, and restrict the range of integration
over $z_r$ to a finite region in order to eliminate the tail contributions.
If we can find a nonzero lower bound for the DOS in this way, 
we have established it independently of the specific decay (e. g.
power law or exponential) at large values of the disorder variable.
First we choose a site $r_0\in S$ for which we restrict the $z_r$
integration to the interval $[-a,a]$. For the remaining integrations on $S$
we define
\begin{equation}
z_r=z_{r_0}+\delta z_r\ \ \ \ {\rm with}\ \delta z_r\in [0,\delta].
\label{int00}
\end{equation}
The above choice of the range of integration on $S$ is sufficient 
but not necessary, i.e. different choices
can be made as long as the range of integration is finite and certain
conditions discussed below are satisfied. In the case considered
$a$ must be chosen large enough to include all singularities of the Green
function. This is the case if it satisfies the inequality
\begin{equation}
0<a-\delta-4(1+\Delta)
\label{conda}
\end{equation}
as we will see below.
Using the notation
$\langle ...\rangle_S'$ for this restricted averaging on $S$ we have
\begin{equation}
\langle N_r\rangle\ge\langle N_r\rangle_S'
\label{ineq00}
\end{equation}
because $N_r$ is nonnegative. Then we can write with 
Eq.(\ref{localdos1})
\begin{eqnarray}
\Big\langle\sum_{r\in S}N_r\Big\rangle_S'
={i\over4\pi}\Big\langle\sum_{r\in S}{\partial\over\partial
z_r}\log\det({\bf 1}+A)\Big\rangle_S'
\nonumber\\
={i\over4\pi}\Big\langle\int_{-a}^adz_{r_0}P(z_{r_0})
(\prod_{r\in S,r\ne r_0}
\int_{z_{r_0}}^{z_{r_0}+\delta}dz_rP(z_r))
{\partial\over\partial z_{r_0}}
\log\det({\bf 1}+A)
\Big\rangle_{\Lambda\backslash S},
\label{inequ1}
\end{eqnarray}
where $\langle ...\rangle_{\Lambda\backslash S}$ refers to the 
(unrestricted, i.e. $z_r \in [-\infty,\infty]$)
averaging over $z_r$ on all lattice sites on $\Lambda$ except the ones on
$S$.\\

We now bound the r.h.s. of Eq. (\ref{inequ1}) from below
by pulling out certain infima of the distribution. 
As a first step, we pull out the infimum of the distribution on $z_{r_0}$. 
This leads to
\begin{equation}
\ge \inf_{z_{r_0}\in [-a,a]}P(z_{r_0})
{i\over4\pi}\Big\langle\int_{-a}^adz_{r_0}
(\prod_{r\in S,r\ne r_0}\int_{z_{r_0}}^{z_{r_0}
+\delta}dz_rP(z_r))
{\partial\over\partial z_{r_0}}\log\det({\bf 1}+A)
\Big\rangle_{\Lambda\backslash S}.
\end{equation}
Furthermore, by pulling out the infimum of the integrand of 
$\prod_{r\in S,r\ne r_0}\int_{z_{r_0}}^{z_{r_0}+\delta}dz_rP(z_r)$
we obtain
\begin{equation}
\ge\inf_{z_{r_0}\in [-a,a]}P(z_{r_0}){i\over4\pi}\Big\langle
\int_{-a}^adz_{r_0}\inf_{z_r\in [z_0,z_0+\delta]}
{\partial\over\partial z_{r_0}}\log\det({\bf 1}+A)
\big[\prod_{r\in S,r\ne r_0}\int_{z_{r_0}}^{z_{r_0}+\delta}P(z_r)dz_r
\big]\Big\rangle_{\Lambda\backslash S}.
\label{5}
\end{equation}
Pulling out the integral over $z_r$ ($r_0\ne r\in S$) eventually yields
\begin{eqnarray}
\ge\inf_{z_{r_0}\in [-a,a]}P(z_{r_0})
\inf_{z_{r_0}\in [-a,a]}
\Big(\int_{z_{r_0}}^{z_{r_0}+\delta}P(z_r)dz_r\Big)^{|S|-1}
\nonumber\\
{i\over4\pi}\Big\langle
\inf_{z_r\in [z_0,z_0+\delta]}\int_{-a}^adz_{r_0}
{\partial\over\partial z_{r_0}}\log\det({\bf 1}+A)
\Big\rangle_{\Lambda\backslash S}
\label{6}
\end{eqnarray}
The disorder distribution on the block $S$ is now taken care of
by the coefficient
\begin{equation}
P_S:=\inf_{z_{r_0}\in [-a,a]}P(z_{r_0})\inf_{z_{r_0}\in [-a,a]}
\Big(\int_{z_{r_0}}^{z_{r_0}+\delta}P(z_r)dz_r\Big)^{|S|-1}
\label{coeff0}
\end{equation}
which multiplies the remaining disorder average over
$\Lambda\backslash S$. $P_S$ is nonzero as long
as the disorder distribution $P(z_r)$ is not vanishing  
in the restricted range of integration. This is certainly true for
unbounded distributions, like a Gaussian. However, compact 
distributions with a narrow range of disorder
will fail to provide a nonzero lower bound. This will be discussed in more
detail below.\\

\subsection{General Lower Bound for the DOS}
Combining Eqs. (\ref{inequ1}), (\ref{6}) and (\ref{coeff0}) we obtain
\begin{equation}
\Big\langle\sum_{r\in S}N_r\Big\rangle_S' \geq
{i\over4\pi}P_S\Big\langle\inf_{z_r\in [z_0,z_0+\delta]}
\Big[\log\det({\bf 1}+A)\Big|_{z_{r_0}=a}-
\log\det({\bf 1}+A)\Big|_{z_{r_0}=-a}\Big]
\Big\rangle_{\Lambda\backslash S}.
\label{7}
\end{equation}
In the next step we isolate the lattice block $S$ from the rest of 
the lattice $\Lambda$
by sending $z_r$ to $\pm\infty$ on the boundary $\partial S$ of $S$. 
(Particles trying to occupy sites on the boundary will either be
trapped or repelled by an infinitely strong barrier).
The boundary $\partial S$ is defined by all sites of $\Lambda$ 
which are not in $S$ but connected with $S$ by the matrix ${\tilde H}$, 
i.e, all sites 
$r\notin S$ with $|r-r'|=1,2$ for any $r'\in S$. (Observe that due to the
definition of the Laplacian the "boundary" is actually two layers around the
block $S$.)
With the above definition of the boundary $\partial S$
the matrix $(1-I_{\partial S}){\tilde H}(1-I_{\partial S})$ separates into
one block
matrix on $S$ and another one on $\Lambda\backslash S\cup\partial S$
\begin{equation}
(1-I_{\partial S}){\tilde H}(1-I_{\partial S})=I_S{\tilde H}I_S+
I_{\Lambda\backslash S\cup\partial S}{\tilde H}
I_{\Lambda\backslash S\cup\partial S}.
\end{equation}
Applying the identities of Appendix A, it follows
that the inverse of 
$(1-I_{\partial S}){\tilde H}(1-I_{\partial S})$ separates into two block
matrices. Consequently,
$\lim_{z_r\to -\infty (r\in\partial S)}A=A_{\Lambda\backslash\partial S}
:=-2i\epsilon 
\big({\tilde H}+i\epsilon\sigma_0\big)^{-1}_{\Lambda\backslash\partial S}$
also separates into two block matrices
\begin{equation}
A_{\Lambda\backslash\partial S}=A_S+A_{\Lambda\backslash S\cup\partial S}.
\end{equation}
Performing the limiting process for the lower bound of the DOS we can use the
fact that $i\log\det({\bf 1}+A)$ is a nondecreasing function of $z_r$ 
\begin{equation}
{\partial\over\partial z_r}{i\over4\pi}\log\det({\bf 1}+A)=N_r\ge0.
\end{equation}
This implies a lower bound for the r.h.s. of Eq.(\ref{7}) if we decrease
the first term in Eq.(\ref{7}) by taking $z_r\to -\infty$ and increase the
second term by taking $z_r\to \infty$ (on the boundary
of $S$). The result of this procedure is the lower bound
\begin{eqnarray}
{i\over4\pi}P_S\Big\langle\inf_{z_r\in [z_0,z_0+\delta]}
\Big[\lim_{z_r\to-\infty (r\in\partial S)}\log\det({\bf 1}+A)
\Big|_{z_{r_0}=a}
\nonumber\\
-\lim_{z_r\to\infty (r\in\partial S)}\log\det({\bf 1}+A)\Big|_{z_{r_0}=-a}
\Big]\Big\rangle_{\Lambda\backslash (S\cup\partial S)}.
\label{8}
\end{eqnarray}
Next, we rewrite the second (negative) term by applying successively Eq. 
(\ref{integral}) for all $r\in\partial S$. This yields
\begin{equation}
\lim_{z_r\to\infty (r\in\partial S)}i\log\det({\bf 1}+A)=4\pi|\partial S|
+\lim_{z_r\to-\infty (r\in\partial S)}i\log\det({\bf 1}+A).
\end{equation}
We therefore have for the expression (\ref{8}) 
\begin{eqnarray}
{i\over4\pi}P_S \Big\langle\inf_{z_r\in [z_0,z_0+\delta]}
\Big[\lim_{z_r\to-\infty (r\in\partial S)}\log\det({\bf 1}+A)
\Big|_{z_{r_0}=a}
\nonumber\\
-\lim_{z_r\to-\infty (r\in\partial S)}\log\det({\bf 1}+A)\Big|_{z_{r_0}
=-a}\Big]\Big\rangle_{\Lambda\backslash (S\cup\partial S)}-P_S|\partial S|.
\end{eqnarray}
There is no contribution from the matrix $A$ on $\Lambda\backslash S\cup
\partial S$, since this
matrix part does not depend on $z_{r_0}=\pm a$. Consequently,
the difference of these contributions gives zero, and
we find a lower bound of the form
\begin{equation}
\Big\langle\sum_{r\in S}N_r\Big\rangle_S'
\ge{i\over4\pi}P_S \inf_{z_r\in [z_0,z_0+\delta]}
\Big[\log\det({\bf 1}+A_S)\Big|_{z_{r_0}=a}
-\log\det({\bf 1}+A_S)\Big|_{z_{r_0}=-a}\Big]-P_S|\partial S|.
\label{9}
\end{equation}
The right hand side of Eq. (\ref{9}) is a difference between a contribution
from the block $S$ (the logarithmic terms) and a boundary contribution
(the $|\partial S|$  term). If the contribution of the block grows with its
volume $|S|$ we find for sufficiently large lattice blocks a positive
lower bound for the r.h.s. of Eq. (\ref{9}). We show below that this
is indeed the case for the considered model of a d--wave SC. 

\subsection{Lower Bound for a 2D d-wave Superconductor}

The growth of the block contribution with the volume $|S|$
follows from the range of the disorder integration on $S$ (Eq. (\ref{int00})
and (\ref{conda})). 
To see this we define
\begin{equation}
H'=I_S{\tilde H}I_S+z_{r_0}\sigma_0 I_S.
\end{equation}
$I_S{\tilde H}I_S$ and $H'$ can be diagonalized by unitary transformations.
An eigenvalue $\lambda_j$ of $I_S{\tilde H}I_S$ satisfies 
$-z_{r_0}+\min \lambda'_j\le\lambda_j\le -z_{r_0}
+\max \lambda'_j$. This implies for the terms in (\ref{9}), where 
$z_{r_0}=\pm a$
\begin{equation}
\mp a+\min \lambda'_j\le\lambda_j\le\mp a+\max \lambda'_j.
\end{equation}
An upper bound of ${\lambda'_j}^2$ can be derived from the eigenvalues of 
$I_S(H+z_{r_0}D\sigma_3)I_S$ (see Appendix B).
It yields $|\lambda'_j|\le 4(1+\Delta)+\delta$, since the deterministic part 
of the Hamiltonian $-\nabla^2\sigma_3+{\hat\Delta}^d\sigma_1$ has an upper
bound $4(1+\Delta)$, and the random
part comes from $\delta z_r$ ($0\le\delta z_r\le\delta$). Thus we obtain
\begin{equation}
-a-4(1+\Delta)\le\lambda_j\le -a+4(1+\Delta)+\delta\ \ \ (z_{r_0}=a)
\label{cond1}
\end{equation}
\begin{equation}
a-4(1+\Delta)\le\lambda_j\le a+4(1+\Delta)+\delta\ \ \ (z_{r_0}=-a).
\label{cond2}
\end{equation}
The condition for $a$ in (\ref{conda}) guarantees that for $z_{r_0}=a$
($z_{r_0}=-a$) all eigenvalues $\lambda_j$ are negative (positive).
Consequently, the argument of the logarithm  for any eigenvalue
$\lambda_j$, 
$1-2i\epsilon/(\lambda_j+i\epsilon)$, is  
$1+i\epsilon$ ~($1-i\epsilon$)
for the first (second) term in  Eq.(\ref{9}).
In order to deal with the branch cut of the
complex logarithm we let $-\epsilon \rightarrow 2 \pi -\epsilon$ for the 
second term in (\ref{9}). Now we can safely let $\epsilon\rightarrow 0$
in both terms and obtain for
\begin{equation}
i\log\det({\bf 1}+A_S)\Big|_{z_{r_0}=a}
-i\log\det({\bf 1}+A_S)\Big|_{z_{r_0}=-a}
\end{equation}
a contribution of $2\pi$ for each of the $2|S|$ eigenvalues $\lambda_j$,
i.e. a total of $4\pi |S|$.
From Eqs. (\ref{ineq00}) and (\ref{9}) it therefore follows that the DOS is 
given by
\begin{equation}
\Big\langle\sum_{r\in S}N_r\Big\rangle \ge P_S(|S|-|\partial S|).
\end{equation}
The average DOS is the sum of the local average DOS, normalized by the
lattice size $|\Lambda|$. Dividing the lattice $\Lambda$ into identical
blocks $S$ we sum over all blocks and obtain after normalization
\begin{equation}
{1\over|\Lambda|}\Big\langle
\sum_{r\in \Lambda}N_r\Big\rangle \ge P_S(1-|\partial S|/|S|).
\label{lowbound}
\end{equation}
Since the lattice block size $|S|$ grows faster than the size of its
boundary $|\partial S|$, there is a finite size which gives a positive bound
on the r.h.s. and therefore a positive lower bound on the DOS.\\

Eq. (\ref{lowbound}) holds for our lattice model of a d--wave SC, given
by the Hamiltonian Eq. (\ref{hamil}) for all unbounded and symmetric
disorder distribution that vanish at large disorder parameters $z_r$.
In particular, the lower bound holds for both power law (e.g. Lorentzian)
and exponential (e.g. Gaussian) distributions. It also holds for
compact distributions of sufficient width, with the width
being determined by the requirement
that the factor $P_S$ must be nonzero when $a$ is chosen according
to the condition Eq.(\ref{conda}) in order to let the DOS on $S$ grow 
with $|S|$. This does not imply that narrow compact distributions
will have a vanishing DOS at the Fermi level.
However, to show the finiteness of the DOS for such distributions 
a more sophisticated method is required.

\section{Conclusions}
In conclusion, we have shown that for rather generic conditions a nonzero
lower bound for the Fermi level density of quasiparticle states exists.
The bound does not depend on the specifics of the "tails" of 
the distribution as both Lorentzian and Gaussian distributions yield 
a nonzero lower bound. This proves that our exact result for the 
case of Lorentzian disorder \cite{zhh} is generic.\\

This result applies to a class of Hamiltonians describing 
2D superconductors with {\it nonlocal} order parameters, like
extended s--wave, p--wave and d--wave SC's. In contrast, for a {\it 
local} isotropic s--wave SC our method will yield  a  vanishing
lower bound, in complete agreement with Anderson's theorem for
nonmagnetic disorder in SC's with a finite order parameter everywhere
on the Fermi surface. It should be noted that our results imply that
the selfconsistent t--matrix approximation \cite{hps} gives 
qualitatively correct physics as long as only the DOS at the Fermi level
is concerned (i.e. for thermodynamic properties). Whether this also holds 
for the dynamic (transport) properties is an interesting
question to be resolved.\\

\noindent
Acknowledgements
 
Two of the authors thank the National Science Foundation for
support under NSF DMR- 94--06678 and 93--57199 (M.H.H.) and 
NSF- DMR-96--00105 (P.J.H.).

\begin{appendix}
\section{Projections of the Green function}
Consider a general square matrix $H$ defined on a lattice $\Lambda$.
$R$ is a subset of $\Lambda$, and $I_R$ is the projector on the 
region $R$ which can be written as a diagonal matrix
\begin{equation}
I_{R,q,q'}=I_{R,q}\delta_{qq'}\ \ \ {\rm with}\ I_{R,q}=\cases{
1&if $q\in R$\cr
0& otherwise\cr
}.
\end{equation}
If the inverse of $H$ and $H+I_RCI_R$ exist then we find the identity
\begin{equation}
(H+I_RCI_R)^{-1}=H^{-1}-H^{-1}({\bf 1}+I_RCI_RH^{-1})^{-1}_RI_RCI_RH^{-1},
\label{ident}
\end{equation}
where
\begin{equation}
(...)^{-1}_R=I_R(I_R...I_R)^{-1}I_R
\label{ident1}
\end{equation}
is the inverse with respect to $R$. From Eq. (\ref{ident}) follows
immediately
\begin{equation}
(H+I_RCI_R)^{-1}=H^{-1}+H^{-1}\Big\{(H^{-1})^{-1}_R\Big[(H^{-1})^{-1}_R
+I_RCI_R\Big]^{-1}_R(H^{-1})^{-1}_R-(H^{-1})^{-1}_R\Big\}H^{-1},
\label{decomp}
\end{equation}
and on $R$ follows
\begin{equation}
I_R(H+I_RCI_R)^{-1}I_R=\Big[(H^{-1})^{-1}_R+I_RCI_R\Big]^{-1}_R
\end{equation}
by means of (\ref{ident1}).
If we choose $C=z_{r_0}\sigma_0$ and let  $z_{r_0}\to\pm\infty$
we obtain with (A2)
\begin{equation}
\lim_{z_{r_0}\to\pm\infty}(H+I_RCI_R)^{-1}=H^{-1}-H^{-1}(H^{-1})^{-1}_R
H^{-1}.
\end{equation}
All matrix elements on $R$ are zero. Therefore, we can write this
expression also as a projection onto $\Lambda\backslash R$ which can
eventually be rewritten as the inverse on $\Lambda\backslash R$
\begin{equation}
\equiv (1-I_R)H^{-1}(1-I_R)-(1-I_R)H^{-1}I_R(H^{-1})^{-1}_RI_RH^{-1}
(1-I_R)=(H)^{-1}_{\Lambda\backslash R}.
\label{decomp1}
\end{equation} 
We use the above identity in the text
with the choice $R = \partial S$, the boundary of the block $S$.\\

\section{Estimation of the eigenvalues}
$H$ and ${\tilde H}=HD\sigma_3$ are Hermitean matrices. Therefore, both 
matrices
can be diagonalized by unitary transformations $U$ and ${\tilde U}$,
respectively.
There are eigenvalues $\lambda_j$ and ${\tilde\lambda}_j$ with
\begin{equation}
\lambda_j=(U H U^\dagger)_{jj}\ \ {\rm and}\ 
{\tilde\lambda}_j=({\tilde U}{\tilde H}{\tilde U}^\dagger)_{jj}.
\end{equation}
Then we have
\begin{equation}
{\tilde\lambda}_j^2=\Big(({\tilde U}{\tilde H}{\tilde U}^\dagger)_{jj}
\Big)^2=({\tilde U}{\tilde H}{\tilde U}^\dagger{\tilde U}{\tilde H}
{\tilde U}^\dagger )_{jj}
=({\tilde U}HD\sigma_3HD\sigma_3{\tilde U}^\dagger)_{jj}.
\end{equation}
Since $H$ and $D\sigma_3$ commute and $(D\sigma_3)^2={\bf 1}$, we obtain for
the r.h.s.
\begin{equation}
({\tilde U}H^2{\tilde U}^\dagger)_{jj}\le\max \lambda_j^2.
\end{equation}
This estimation holds for any projection of $H$ and ${\tilde H}$ 
on a region $S$ as long
as the relation ${\tilde H}=HD\sigma_3$ is valid on $S$.
We apply the above inequality in our estimation
of the eigenvalues of the projection of $\tilde{H}$ on the lattice block
$S$.
\end{appendix}
\end{document}